\documentclass[pra,twocolumn,showpacs]{revtex4}
\usepackage{exscale}
\usepackage{amsmath}
\usepackage[cspex, bbgreekl]{mathbbol}
\usepackage{amscd}
\usepackage{amsfonts}
\usepackage{amsgen}
\usepackage{amssymb}
\usepackage{amstext}
\usepackage{amsxtra}
\usepackage{mathrsfs}
\usepackage{graphicx} 
\usepackage{epsfig} 
\usepackage{amssymb}
\usepackage{amsmath}
\usepackage{amsfonts}
\usepackage{ifthen} 
\usepackage{fancyhdr}
\usepackage{graphics,color,dsfont}
\usepackage{epic,rotating}
\usepackage{pstricks}
\usepackage{color}

\newcommand{\diff}{\mathrm{d}}
\newcommand{\eh}{\mathrm{e}}

\newcommand{\rb}{\protect{\mathbf{r}}}
\newcommand{\kb}{\protect{\mathbf{k}}}
\newcommand{\Br}{\protect{\mathrm{B}}}

\begin{document}

%
\title{Interaction of Impurity Atoms in Bose-Einstein-Condensates}
%
\author{Alexander Klein}
\author{Michael Fleischhauer}
\affiliation{Fachbereich Physik, Technische Universit{\"a}t Kaiserslautern, D-67663 Kaiserslautern, Germany}
\begin{abstract}
The interaction of two spatially separated impurity atoms 
through phonon exchange in a Bose-Einstein condensate is studied within 
a Bogoliubov approach. The impurity atoms are held by deep and narrow
trap potentials and experience level shifts which 
consist of a mean-field part and vacuum contributions 
from the Bogoliubov-phonons. In addition there is a conditional
energy shift resulting from the exchange of phonons between the impurity 
atoms. 
\end{abstract}
\maketitle


\section{Introduction}


The ability to engineer the collisional interaction of
ultra-cold individual atoms or ions  as well as
degenerate ensembles of atoms, such as Bose-Einstein condensates (BECs)
\cite{BEC}
has dramatically improved in the last couple of years
by the development of quantum-optical tools such as single-atom micro-traps
\cite{Grangier-Nature-2001,Raizen-PRL-2002,Ertmer -PRL-2002}, 
optical lattices 
\cite{Kasevich-Science-2002,Bloch-Nature-2002,Bloch-Nature-2003}, atom-chips
\cite{atom-chips} and others. Controlled collisional 
interactions of individual atoms are of fundamental interest but also
have important potential applications in quantum information processing
\cite{quantum-information}. Recently the coupling of single-atom 
quantum dots to Bose-Einstein condensates was studied in
\cite{Recati-condmat-2004} and 
the use of an impurity atom in
a 1-dimensional optical lattice as an atom transistor was proposed 
\cite{Micheli-quantph-2004}.
We here study the mutual interaction between two separated, well localized 
impurity atoms through the exchange of Bogoliubov phonons 
in a BEC at zero temperature. When the impurity atoms undergo a state-dependent
scattering with the condensate atoms, in addition to mean-field level shifts
and levels shifts from the interaction with the vaccum fluctuations
of the Bogoliubov phonons a conditional level shift emerges which results 
from phonon exchange between the impurities. This conditional shift 
is calculated and its dependence on trap geometry,
impurity separation and the strength of the interactions within the
condensate is studied. 

In section II we derive an effective coarse-grained 
interaction hamiltonian for the impurity
atoms and relate the level shifts to correlation functions of 
quasi-particle excitations. These will then be calculated 
within a Bogoliubov approximation for a condensate in a box potential
in section III. It is shown that the coupling between the impurity atoms 
is strongest for a highly asymmetric geometry. For this reason we consider
in section IV a quasi-one dimensional condensate. A simple analytic 
expression for the level shift is derived using a Thomas-Fermi approximation.


\section{Effective interaction of impurity atoms in a BEC}

\begin{figure}[ht]
\includegraphics[width=7.5cm]{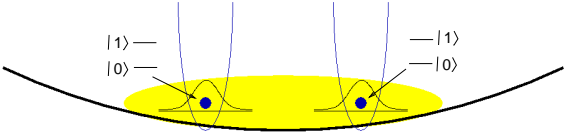} 
\caption{Impurity atoms held by tight confining potentials
in a Bose-Einstein condensate. When in internal state $|1\rangle$ the
atoms undergo s-wave scattering interactions with the condensate.
}
\label{fig1}
\end{figure}

We here consider a Bose-Einstein condensate at $T=0$ with 
impurity atoms at fixed locations, which can be realized e.g.
by tightly confining trap potentials as shown in Fig.\ref{fig1}. 
The traps are seperated such that any direct interaction of the atoms 
can be excluded.
The atoms are assumed to have
two relevant internal states $|0\rangle$ and $|1\rangle$ and
shall undergo s-wave scattering interactions 
with the atoms of the BEC if they are
in state $|1\rangle$. If the traps are sufficiently deep, 
the atoms will stay in the
corresponding ground state $\phi_0$. In this case the  
interaction hamiltonian of the condensate and the impurities has the form
\begin{equation}\label{HWWdef}
  \begin{split}
  \hat{H}_\mathrm{int} =&\sum_{\alpha,\beta} \left|\alpha,\beta\right
\rangle\left\langle \alpha,\beta\right| \\
  &\left( \frac{\kappa_\alpha}{2} \int \diff \mathbf{r} \left|\phi_0
(\mathbf{r}-\mathbf{r}_1)\right|^2 \hat{\psi}^\dagger (\mathbf{r})
\hat{\psi}(\mathbf{r}) \right.\\
  &\left.+ \frac{\kappa_\beta}{2} \int \diff \mathbf{r} \left|
\phi_0(\mathbf{r}-\mathbf{r}_2)\right|^2 \hat{\psi}^\dagger 
(\mathbf{r})\hat{\psi}(\mathbf{r}) \right) \,,
  \end{split}
\end{equation}
where $\left|\alpha,\beta\right\rangle$ denotes the $\alpha$-th internal 
state of the first and the $\beta$-th internal state of the second impurity 
atom. The coupling to the condensate is described by the state dependent 
coupling constant $\kappa_\alpha$ with $\kappa_0=0$ and $\kappa_1=\kappa$.
 The condensate wave-function is denoted by $\hat\psi$. The ground state 
function of the impurities is given by
\begin{equation}
  \phi_0(\mathbf{r})=\frac{1}{ \sqrt{\sqrt{\pi^{3}} z_0^3}} 
\exp\left( -\frac{r^2}{2z_0^2}\right) 
\end{equation}
with $z_0=\sqrt{\hbar/m_\mathrm{S}\omega_0}$, $m_\mathrm{S}$ being the 
mass of the impurity atoms and $\omega_0$ the frequency of the confining traps.


In order to derive an effective Hamiltonian for the two impurity atoms 
it is convenient to first separate the interaction (\ref{HWWdef}) into a
mean-field and a fluctuation part
\begin{equation}\label{HWWdef2}
  \begin{split}
  \hat{H}_\mathrm{int} =&
\left|1\right\rangle_{11}\!\left\langle 1\right|
\frac{\kappa}{2}
\left\langle\hat C_1(t)\right\rangle
+ 
\left|1\right\rangle_{22}\!\left\langle 1\right|
\frac{\kappa}{2}
\left\langle\hat C_2(t)\right\rangle
\\
&+ \sum_{\alpha,\beta} 
\left|\alpha,\beta\right\rangle\left\langle \alpha,\beta\right|
\left( \frac{\kappa_\alpha}{2} \hat B_1(t) +\frac{\kappa_\beta}{2}
\hat B_2(t)\right)
\,,
  \end{split}
\end{equation}
where
\begin{equation}
  \hat C_l(t)=\int \! \diff \mathbf{r} \, \left|\phi_0(\mathbf{r}-\mathbf{r}_l)\right|^2 \hat \psi^\dagger(\mathbf{r},t) \hat \psi(\mathbf{r},t) \,,
\end{equation}
and
\begin{equation}
  \hat B_l(t)= \hat C_l(t)-\left\langle \hat C_l(t) \right\rangle\,.
\end{equation}
The terms in the first line of eq.(\ref{HWWdef2}) result in a mean-field
level shift of the internal state $|1\rangle$.
They are of no interest in the present discussion
and will be absorbed in the free Hamiltonian of the impurity atoms. 

We proceed by deriving an equation of motion for the statistical operator 
of the impurities interacting with the BEC. Within the usual Born 
approximation and as outlined in Appendix A one finds 
\begin{eqnarray}
  \partial_t \tilde \varrho_{10,00}
&=&
-\frac{\kappa^2}{4\hbar^2}\int_{t_0}^t \! 
\diff t' \, \tilde \varrho_{10,00}(t') 
\left\langle \tilde B_1(t) \tilde B_1(t')\right\rangle \label{rho1}\\
  \partial_t \tilde \varrho_{01,00} &=& 
-\frac{\kappa^2}{4\hbar^2}\int_{t_0}^t \! \diff t' \, 
\tilde \varrho_{01,00}(t') \left\langle \tilde B_2(t) 
\tilde B_2(t')\right\rangle \label{rho2}\\
\partial_t \tilde \varrho_{11,00} &=& 
-\frac{\kappa^2}{4\hbar^2}\int_{t_0}^t \! \diff t' \, 
\tilde \varrho_{11,00}(t') \Bigl\{ \left\langle \tilde B_1(t) 
\tilde B_1(t')\right\rangle  \label{rho3}\\
 &+& \left\langle \tilde B_1(t) \tilde B_2(t')\right\rangle + \enspace \textrm{terms with}\, \, 1\, 
\leftrightarrow\,  2\enspace \Bigr\} \,.\nonumber
\end{eqnarray}
where the tilde denotes quantities in the interaction 
picture and 
the matrix elements of the statistical operator
are denoted by $\tilde \varrho_{\alpha\beta,\gamma\delta}=
\left\langle \alpha\beta,t \right|\tilde \varrho \left| \gamma\delta,t \right
\rangle $. 
The correlations $\left\langle \tilde B_l\tilde B_{l^\prime}\right\rangle$
are calculated using the standard Bogoliubov approach, i.e. by setting
\begin{equation}
\hat\psi(\mathbf{r},t) = \psi_0(\mathbf{r}) 
+\hat \xi(\mathbf{r},t)
\end{equation}
with $\psi_0$ being the solution of the Gross-Pitaevskii equation
and $\hat\xi$ a small operator-valued correction and neglecting
higher-order terms in $\hat \xi$
 (see Appendix B). 
Within the Bogoliubov approach we disregard 
terms of the order $\mathcal{O}\left(\xi^4\right)$ 
in $\left\langle \tilde B_i\tilde B_j\right\rangle$ and find
\begin{equation} \label{Korrelationsfunktion}
  \left\langle \tilde B_l(t) \tilde B_{l'}(t')\right\rangle = \left.\sum_j\right.^\prime \eh^{- \frac{i}{\hbar}E_j(t-t')}  S_j(l,l').
\end{equation}
The $E_j$'s are the Bogoliubov energies and
\begin{equation} \label{Abkkorr}
\begin{split}
  & S_j(l,l')= \int \!\diff \mathbf{r}\, \left| \phi_0(\mathbf{r}-\mathbf{r}_l)\right|^2 \psi_0(\mathbf{r}) (u_j(\mathbf{r})-v_j(\mathbf{r}))\\
  &\times \int \!\diff \mathbf{r}'\, \left| \phi_0(\mathbf{r}'-\mathbf{r}_{l'}) \right|^2 \psi_0(\mathbf{r}') (u^*_j(\mathbf{r}')-v^*_j(\mathbf{r}')) \,.
\end{split}
\end{equation} 
The functions $u_j$ and $v_j$ are the solutions of the 
Bogoliubov-de Gennes equations (cf. Appendix B) and 
the prime at the sum indicates that the ground state is excluded. 

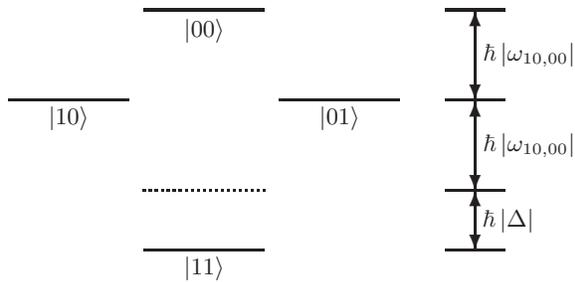
\begin{figure}[ht]
   \setlength{\unitlength}{0.8cm}
  \begin{center}
  \begin{picture}(10,6.5)
    \thicklines
    \put(3,6){\line(1,0){2}}        \put(3,1.45){\makebox(2,0.5){$\left|11\right\rangle$}}
    \put(0.75,4.5){\line(1,0){2}}   \put(0.75,3.95){\makebox(2,0.5){$\left|10\right\rangle$}}
    \put(5.25,4.5){\line(1,0){2}}   \put(5.25,3.95){\makebox(2,0.5){$\left|01\right\rangle$}}
    \put(3,2){\line(1,0){2}}        \put(3,5.4){\makebox(2,0.5){$\left|00\right\rangle$}}
    \dottedline{0.1}(3,3)(5,3)
    \put(8,3){\line(1,0){1}}
    \put(8,4.5){\line(1,0){1}}
    \put(8,6){\line(1,0){1}}
    \put(8,2){\line(1,0){1}}
    \put(8.5,4){\vector(0,-1){1}}
    \put(8.5,3.9){\vector(0,1){0.6}}       
\put(8.7,3.5){\makebox(1.4,0.5){$\hbar\left|\omega_{10,00}\right|$}}
    \put(8.5,5){\vector(0,-1){0.5}}
    \put(8.5,4.9){\vector(0,1){1.1}}         
\put(8.7,5){\makebox(1.4,0.5){$\hbar\left|\omega_{10,00}\right|$}}
    \put(8.5,2.5){\vector(0,-1){0.5}}
    \put(8.5,2.4){\vector(0,1){0.6}}     
\put(8.7,2.25){\makebox(0.7,0.5){$\hbar\left|\Delta\right|$}}
  \end{picture}
  \caption{Energy scheme of the effective Hamiltonian for symmetric 
arrangement of impurity atoms.
Here a negative sign of $\Delta$ was assumed although positive
values are possible.
\label{Energieschema}}
  \end{center}
\end{figure}

A calculation of the correlation functions 
shows that the often used Markov approximation cannot straight-forwardly
applied to eqs.(\ref{rho1}-\ref{rho3}). Instead 
we first use a Laplace transformation. Setting $t_0=0$ we find
\begin{equation} \label{Laptraforho}
  \mathcal{L}\left[\tilde\varrho_{\alpha\beta,\gamma\delta}(t)\right](p)
= \frac{\tilde\varrho_{\alpha\beta,\gamma\delta}(0)}{
p+\frac{1}{4\hbar^2}M_{\alpha\beta,\gamma\delta}(p)}
\end{equation}
with 
\begin{equation}
\begin{split}
 &M_{\alpha\beta,\gamma\delta}(p)= \\
 &\left.\sum_j \right.^{\prime}  \left\{  S_j(1,1)\left( \frac{\kappa^2_\alpha - \kappa_\alpha\kappa_\gamma }{p+\frac{i}{\hbar}E_j}+ \frac{ \kappa_\gamma^2 - \kappa_\alpha\kappa_\gamma}{p-\frac{i}{\hbar}E_j} \right) \right. \\
 & \phantom{\left.\sum_j \right.^{\prime} \{}+ S_j(1,2)\left( \frac{ \kappa_\alpha\kappa_\beta - \kappa_\beta\kappa_\gamma}{p+\frac{i}{\hbar}E_j}+ \frac{\kappa_\gamma\kappa_\delta - \kappa_\beta\kappa_\gamma }{p-\frac{i}{\hbar}E_j} \right)   \\
  &\phantom{\left.\sum_j \right.^{\prime} \{}+  S_j(2,1)\left( \frac{ \kappa_\alpha\kappa_\beta - \kappa_\alpha\kappa_\delta }{p+\frac{i}{\hbar}E_j}+ \frac{\kappa_\gamma\kappa_\delta - \kappa_\alpha\kappa_\delta }{p-\frac{i}{\hbar}E_j} \right)  \\
  &\phantom{\left.\sum_j \right.^{\prime} \{}+\left. S_j(2,2)\left( \frac{ \kappa_\beta^2 - \kappa_\beta\kappa_\delta  }{p+\frac{i}{\hbar}E_j}+ \frac{\kappa_\delta^2 - \kappa_\beta\kappa_\delta  }{p-\frac{i}{\hbar}E_j} \right) \right\} \,.
\end{split}
\end{equation}
In general, the Laplace transformation (\ref{Laptraforho}) 
cannot be inverted analytically. However, if we are interested only 
in a coarse-grained time evolution, it is possible to neglect 
the $p$-dependence of $M_{\alpha\beta,\gamma\delta}$, which amounts to
$M_{\alpha\beta,\gamma\delta}(p) \rightarrow 
M_{\alpha\beta,\gamma\delta}(0)$. In the coarse-grained picture
the interaction of the impurity atoms with the condensate
simply results into level shifts, i.e.
\begin{equation}
  \tilde\varrho_{\alpha\beta,\gamma\delta} (t) =\tilde\varrho_{\alpha\beta,\gamma\delta} (0) \eh^{-i\omega_{\alpha\beta,\gamma\delta}t}.
\end{equation}
The corresponding frequencies read
\begin{equation} \label{Frequenzen}
\begin{split}
  &\omega_{\alpha\beta,\gamma\delta} = \frac{1}{4 \hbar^2 i}
M_{\alpha\beta,\gamma\delta}(0)= \\
  &\frac{1}{4\hbar}\left. \sum_j \right.^\prime \frac{1}{E_j}
\left\{ S_j(1,1) \left( \kappa_\gamma^2 - \kappa_\alpha^2\right) 
+ S_j(2,2)\left(  \kappa_\delta^2 - \kappa_\beta^2\right) \right.\\
  &+\left. \left[ S_j(1,2)+ S_j(2,1)\right]
 \left( \kappa_\gamma \kappa_\delta - \kappa_\alpha\kappa_\beta\right)  \right\} \,.
\end{split}
\end{equation}
This corresponds to an effective - coarse-grained - Hamiltonian 
\begin{equation}
\begin{split}
  \tilde H_{\mathrm{eff}}= &\left| 10\right\rangle \left\langle 10\right| \hbar\omega_{10,00} +  \left| 01\right\rangle \left\langle 01\right| \hbar\omega_{01,00} \\
  &+  \left| 11\right\rangle \left\langle 11\right| \hbar\omega_{11,00}  \,.
\label{Heff}
\end{split}
\end{equation}
The energy scheme of this Hamiltonian 
is shown in figure \ref{Energieschema}. 
One recognizes from (\ref{Heff}) 
for symmetric impurity locations
a level shift 
\begin{eqnarray}
\delta &=& \omega_{10,00}=\omega_{01,00}\nonumber\\
&=&-\frac{\kappa^2}{4\hbar}{\sum_j}^\prime
S_j(l,l) < 0
\end{eqnarray}
of each impurity atom
independent of the presence of the other. This level shift is due to
the interaction
with vacuum fluctuations of the Bogoliubov quasi particles (phonons). 
In addition there is a conditional level shift due to the exchange
of Bogoliubov quasi particles (phonons) between the two impurities:
\begin{eqnarray} 
  \Delta &=& \omega_{11,00}-\omega_{10,00}- \omega_{01,00} \nonumber\\
&=&-\frac{\kappa^2}{4\hbar} \left. \sum_j \right.^\prime \frac{1}{E_j}\left\{ S_j(1,2) + S_j(2,1)\right\} \,.\label{Deltaallg}
\end{eqnarray}%
%
%

It should be noted that the coarse-graining approximation is consitent 
with the collective level shift only if 
\begin{equation}
 \Delta \ll \frac{1}{\hbar}\left.\min_j\right.^\prime \left( E_j\right) 
\end{equation}
where the prime indicates that the ground state is excluded. 
In the following we will explicitly calculate the levels shifts for a 
homogeneous condensate, for an ideal condensate in a harmonic trap 
and a weakly interacting condensate in a trap in the Thomas-Fermi limit.


\section{Homogeneous Condensate}

\begin{figure} [htb]
  \begin{center}
  \scalebox{0.44}{\includegraphics{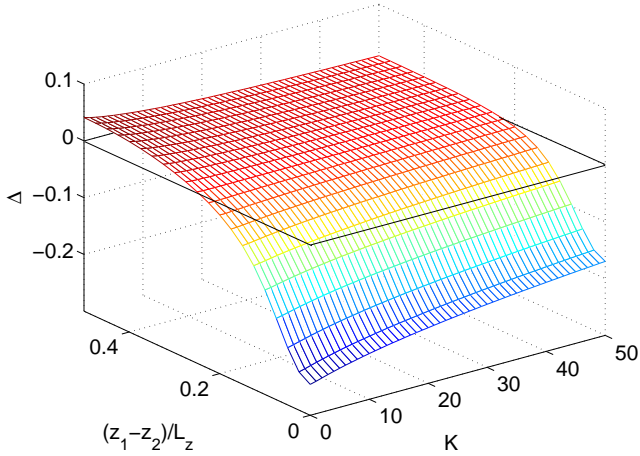}}
  \caption{Energy shift in units of $\kappa^2 N_0 m_\Br 
 L_z^2/\hbar^3 V^2$ for two impurities in a homogeneous 
condensate with periodic boundary conditions. The  
interaction of BEC atoms is characterized by the dimensionless
parameter $K=gN_0 2 m_\Br L_z^2/\hbar^2 V$. 
The impurities are located on the 
$z$-axis,
$L_x=L_y=0.5L_z$, and $z_0=0.05L_z$. \label{VerschPmWW3Dbild}}
  \end{center}
\end{figure}

In this section we calculate the energy shifts $\delta$ 
and  $\Delta$ for the case of an interacting, 
homogeneous condensate with periodic boundary conditions 
of spatial periodicity $L_x$, $L_y$, $L_z$, respectively. 
The solutions of the Bogoliubov-de Gennes equations are then 
given by plane waves 
$u_\kb=(f_\kb^+ + f_\kb^-)/2$ and $v_\kb=(f_\kb^+ - f_\kb^-)/2$ 
with
\begin{equation} \label{fpmloesungen}
  f^\pm_\mathbf{k} (\mathbf{r})= \frac{1}{\sqrt{V}} \left( \sqrt{\frac{E_\mathbf{k}}{\varepsilon^0_\mathbf{k}}}\right)^{\pm 1} \eh^{i \mathbf{k} \cdot \mathbf{r}} \,.
\end{equation}
The wave vectors $\kb$ have to be chosen in such a way that 
they fulfill the periodic boundary conditions. %
Since the $u_\kb$ and $v_\kb$'s have to be orthogonal to the ground state
the case $\mathbf{k}=0$ is excluded. The Bogoliubov energies are given by
\begin{equation} \label{BogolenergienPmWW3D}
  E_\mathbf{k}=\sqrt{\varepsilon^0_\mathbf{k}\left( \varepsilon^0_\mathbf{k} + 2 g n_0 \right)} 
\end{equation}
with $\varepsilon_\mathbf{k}^0=\hbar^2k^2/2m_\Br$. By extending the 
integral in equation (\ref{Abkkorr}) over the whole $\mathbb{R}^3$, which 
is possible due to the effective cut-off provided by the impurity
state wavefunctions $\phi_0$,
 one can easily calculate the correlation functions:
\begin{equation}
\begin{split}
  &\left\langle \tilde B_l(t) \tilde B_{l'}(t')\right\rangle = \frac{N_0}{V^2} \left. \sum_\mathbf{k} \right.^\prime \frac{\varepsilon^0_\mathbf{k}}{E_\mathbf{k}}\exp\left( -\frac{i}{\hbar}E_{\mathbf{k}} (t-t') \right) \\
  &\qquad \times \exp \left(i\mathbf{k}\cdot\left(\mathbf{r}_l-\mathbf{r}_{l'} \right)-\frac{z_0^2\mathbf{k}^2}{2}  \right) \,.
\end{split}
\end{equation}
Here $N_0$ denotes the number of atoms 
in the condensate and 
$V=L_x L_y L_z$. With this one finds 
\begin{eqnarray}
\delta &=&  - \frac{\kappa^2 N_0}{4 \hbar V^2} 
  {\sum_{\mathbf{k}}}^{\prime}  
\frac{\varepsilon_\mathbf{k}^0}{E^2_\mathbf{k}}
\exp\left( -\frac{z_0^2{k}^2}{2} \right) \\
 \Delta &=& - \frac{\kappa^2 N_0}{2 \hbar V^2} 
{\sum_{\mathbf{k}}}^{\prime}  
\frac{\varepsilon_\mathbf{k}^0}{E^2_\mathbf{k}} 
\cos\left(\mathbf{k}\cdot\Delta \rb\right)
  \exp\left( -\frac{z_0^2{k}^2}{2} \right),\,\,
\end{eqnarray}
where $\Delta\rb =\rb_1-\rb_2$.
The sum over the Bogoliubov quasi momenta converges
due to the exponential term which effectively cuts off momenta 
with $k\gg 1/z_0 $. Obviously $2 \delta=\Delta$ for $\mathbf{r}_1=\mathbf{r}_2$
and $|\Delta|\le 2 |\delta|$. 
The conditional energy shift $\Delta$ is shown in 
figure \ref{VerschPmWW3Dbild}.
For very small distances of the impurities
$\Delta$ is negativ and its absolute value approaches its maximum, 
i.e. that of $2 \delta$. 
For increasing distance the value of $\Delta$ increases monotonously
and eventually changes its sign. The monotonous increase would correspond to
an attractive force between the impurity atoms if they could move freely.
 One recognizes, that for 
larger values of the dimensionless interaction parameter
$K\sim g$ the energy shift decreases and the spatial dependence
becomes less pronounced. 
This can be explained by the increasing self-energy 
of the Bogoliubov excitations.

\begin{figure} [htb]
  \begin{center}
\setlength{\unitlength}{1cm}
\scalebox{0.44}{\includegraphics{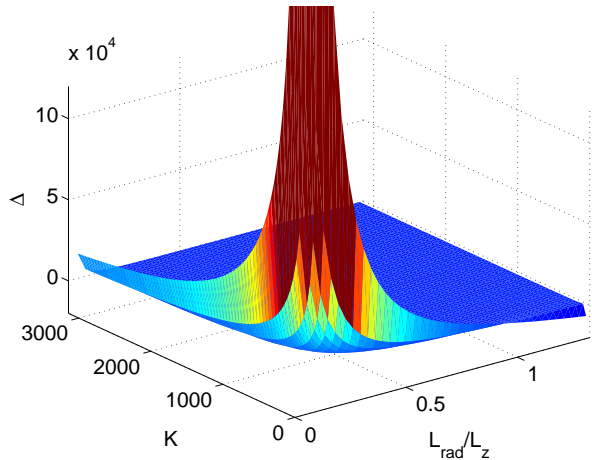}} 
\caption{Influence of condensate geometry on energy shift
in a box with periodic boundary conditions. $\Delta$ is in units of 
$\kappa^2 2 m_\Br/4 \hbar^3$, $K$ is defined as in 
figure \ref{VerschPmWW3Dbild}. 
The impurities are located on the $z$-axis with a
distance of $0.5L_z$. $L_z=12 \cdot 10^{-6}\mathrm{m}$,  and
$L_x=L_y$ has been varied.\label{warum1DPmWW}}
  \end{center}
\end{figure}

It is also very instructive to consider the  dependence of the 
conditional level shift $\Delta$ on the condensate geometry, i.e. 
on the ratio $L_\mathrm{rad}/L_z$, where $L_\mathrm{rad}=L_x=L_y$. 
This is illustrated in figure \ref{warum1DPmWW}. One recognizes that the 
absolute value of the energy shift increases as the ratio 
$L_\mathrm{rad}/L_z$ decreases. Thus the energy shift is largest for
a highly non-symmetric geometry of the BEC. The strongest effect is thus 
to be expected in a quasi one-dimensional condensate. 
For this reason we will investigate in the following section the energy shift 
in the case of a BEC in a harmonic trap only for 
a one-dimensional condensate.

\section{1-D condensate in a trap}

\begin{figure}[htb]
\begin{center}
  \setlength{\unitlength}{1cm}
  \begin{pspicture}(8,7)
    \rput{0}(4,3.8){\scalebox{0.45}{\includegraphics{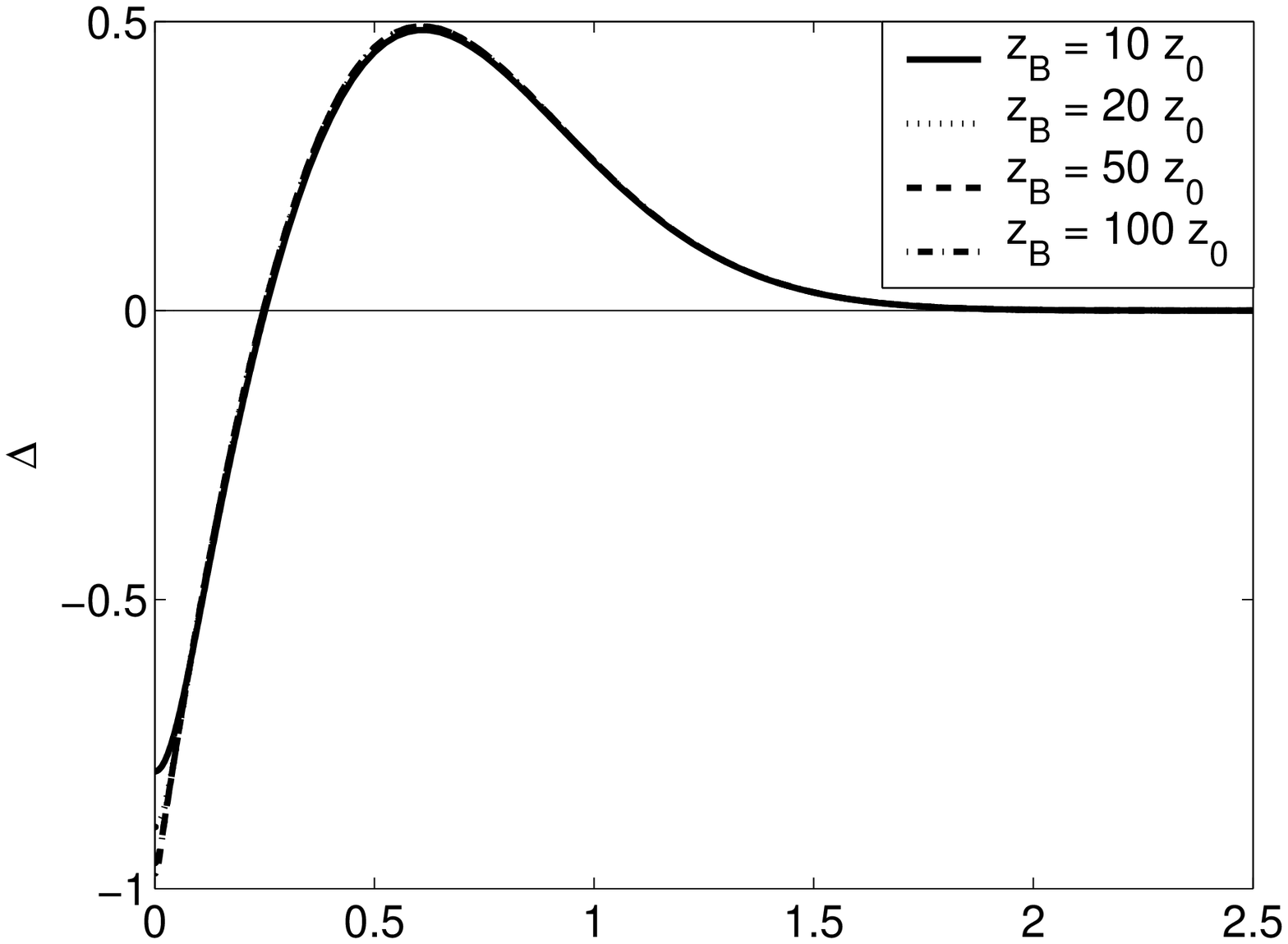}}}
+    \rput{0}(5.4,3.1){\scalebox{0.24 0.22}{
       \includegraphics{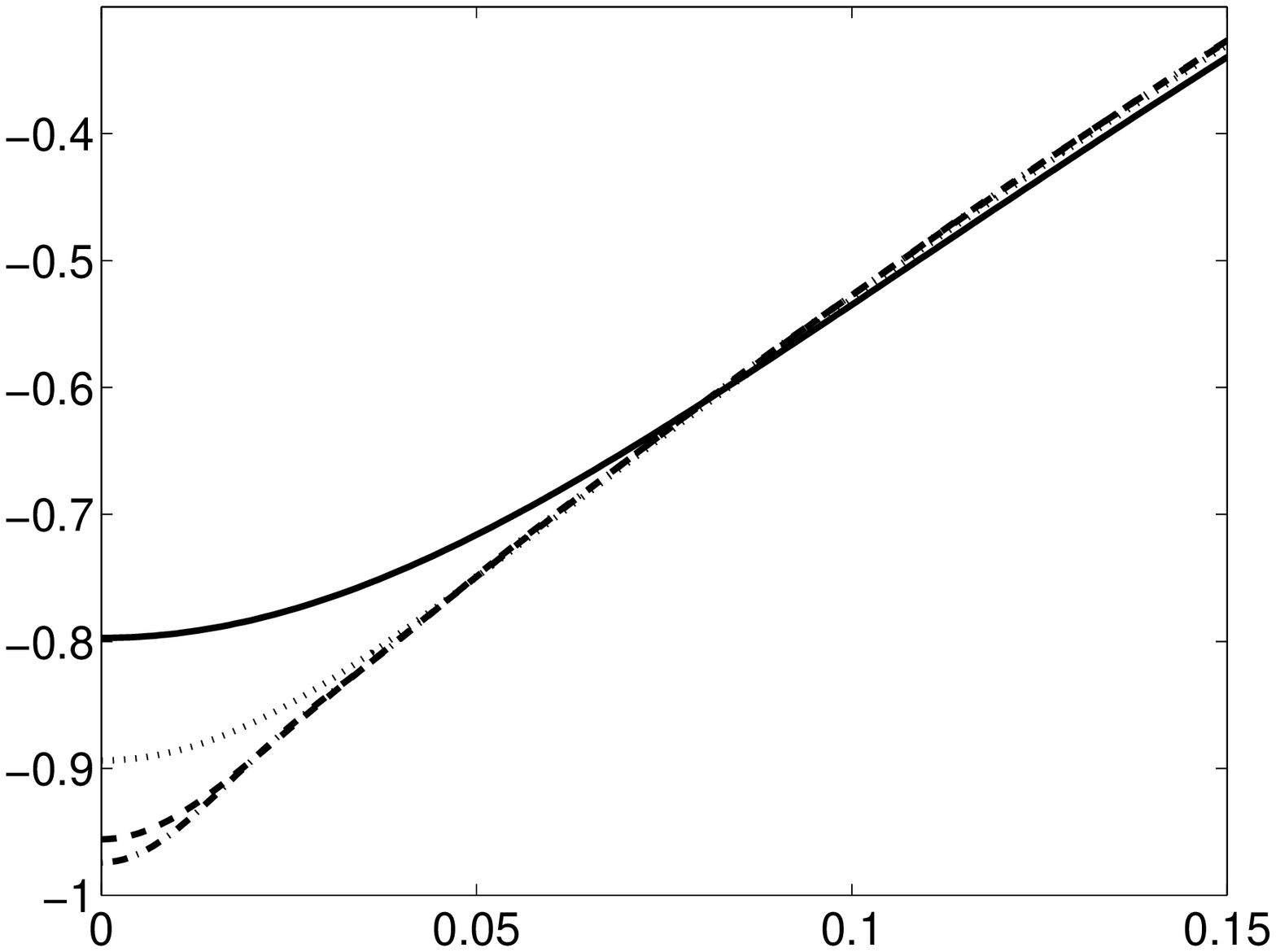}}}
    \rput{0}(4.5,0.4){\makebox(1,0.5){\normalsize $\check{z}_1$}}
  \end{pspicture}
  \caption{Energy shift $\Delta$ in an ideal 1-D condensate 
as function of impurity distance 
for different ratios of $z_\Br/z_0$. 
The shift is given in units of 
$\kappa_{\mathrm{1D}}^2 N_0/2 \hbar^2 \omega_\Br \pi 
\left(z_\Br^2 + z_0^2 \right)$. The inset shows a magnification 
for small distances. Here $\check{z}_1=-\check{z}_2$. \label{HoWW1Dverscheps}}
\end{center}
\end{figure}

In this section we consider a quasi one-dimensional condensate 
confined in an harmonic trap 
$V_\mathrm{ext}= m_\Br \omega_\Br^2 z^2/2$. 
We first consider the case of an ideal, i.e. noninteracting gas.
In this case
the solutions of the Gross-Pitaevskii equation and the Bogoliubov-de 
Gennes equations are just the solutions of the harmonic oscillator:
\begin{gather}
  \psi_0(z) = \sqrt{\frac{N_0}{\sqrt{\pi}z_\Br}} 
\eh^{-{z^2}/{2z_\Br^2}} \\
  u_j(z)=\frac{1}{\sqrt{2^j j! \sqrt{\pi} z_\mathrm{B}}} 
\eh^{-{z}^2/2 z_\mathrm{B}^2}
 H_{j} \left( \frac{z}{z_\mathrm{B}}\right) \\
  v_j(z)=0 \\
  E_j = j \hbar \omega_\Br \,.
\end{gather}
Here $z_\Br=\sqrt{\hbar/m\omega_\Br}$ is the ground-state width of the 1-D 
harmonic trap.
By calculating the integrals in equation (\ref{Abkkorr}) one finds
\begin{equation}
\begin{split}
  \Delta =- \frac{\kappa_\mathrm{1D}^2}{2 \hbar}  
\sum_{\nu =1}^\infty &\frac{1}{\hbar\nu \omega_\Br} 
\frac{N_0 \exp\left(-\check{z}_1^2 -\check{z}_{2}^2\right)}
{\pi (z_0^2 + z_\mathrm{B}^2)}  \frac{\check{z}^\nu}{2^{\nu}  \nu!} \\
  & \times H_\nu \left( \check{z}_{1}\right)  H_\nu \left( \check{z}_{2}\right) \,,
\end{split}
\end{equation}
where $\check{z}=z_\Br^2/(z_\Br^2+z_0^2)$ and 
$\check{z}_l = {z_l}/{\sqrt{z_\Br^2+z_0^2}} $. We also have introduced the
one-dimensional coupling constant 
$\kappa_\mathrm{1D}=\kappa/(2\pi a_\bot^2)$ with the radial 
confinement $a_\bot^2=\hbar/m_\Br \omega_\bot$. 
The conditional level shift is shown 
in figure \ref{HoWW1Dverscheps} for different widths $z_0$ 
of the impurity traps. As expected the shape of the curves 
coincides for distances larger than the ground state width of the 
impurity traps. Distances smaller than $z_0$ are excluded 
because we have assumed that there is no direct scattering 
interaction between the impurity atoms.

It is interesting to note that different from the case of a condensate in a 
box, the force between the impurities is not always attractive.
One recognizes that this is only the case if the distance is 
sufficiently small. If the distance is larger than a certain value, 
in our case $\check{z}_1-\check{z}_2\approx 2\cdot 0.6$, the 
force becomes  repulsive. 

We now consider the case of a weakly interacting 1-D gas.
In order to solve the Gross-Pitaevskii equation we make use of 
the Thomas-Fermi (TF) approximation. Although the results obtained in this way 
cannot be smoothly connected to the ideal-condensate case, the TF approximation
allows to derive a compact expression for the level shift.
The TF condensate wavefunction is given by 
\begin{equation}
  \psi_0(z)=\sqrt{\frac{\mu}{g_\mathrm{1D}}\left(1-\frac{z^2}{R_\mathrm{TF}^2} \right)} \,,
\end{equation}
where the TF radius is given by 
$R_\mathrm{TF}=\sqrt{2\mu/m_\Br \omega_\Br^2}$. $\mu$ denotes 
the chemical potential and  
the one dimensional interaction parameter $g_\mathrm{1D}$ is defined analogous to $\kappa_\mathrm{1D}$.
To solve the Bogoliubov-de Gennes equations analytically 
further approximtions are needed as discussed in \cite{Oehberg-PRA-1997}. 
We here take over the results for the functions $f^\pm_j$ 
obtained in \cite{Petrov-PRL-2000}:
\begin{equation} \label{LoesungPetrov}
  f^\pm_j (z)=\sqrt{\frac{2j+1}{2R_\mathrm{TF}}} \left[ \frac{2\mu}{E_j}\left(1-\frac{z^2}{R_\mathrm{TF}^2} \right)\right]^{\pm \frac{1}{2}} P_j\left(\frac{z}{R_\mathrm{TF}} \right) 
\end{equation}
with the energies $ E_j = \hbar \omega_\Br \sqrt{j(j+1)/2}$. 
The $P_j$ are Legendre polynomials.
Using the completeness of the Legendre polynomials one can explicitely 
evaluate expression (\ref{Deltaallg}). One finds that 
the sum including the $j=0$ term vanishes if the overlap of the 
impurity wavefunctions is negligible:
\begin{equation}
\begin{split} \label{HmWW1Dfast0}
&\sum_{j=0}^\infty \frac{S_{j}(1,2)}{E_{j}} \\
&\quad=   -\frac{1}{2 g_\mathrm{1D}} \int \! \diff z \, \left| \phi_0(z-z_1)\right|^2 \left| \phi_0(z-z_2)\right|^2 \approx 0 \,.
\end{split}
\end{equation}
Thus the energy shift (\ref{Deltaallg}) 
is determined only by the $j=0$ term which 
yields the simple expression :
\begin{equation} \label{DeltaHmWW1D}
  \Delta = \frac{\kappa_\mathrm{1D}^2}{8 \hbar R_\mathrm{TF}g_\mathrm{1D}} \,.
\end{equation}
This result does not depend on the distance of the impurity atoms which is
due to the Thomas-Fermi approximation. The shift is always positive and 
becomes larger for smaller interactions in the BEC and for larger 
1D confinement. It is instructive to express $\Delta$ in terms of
the impurity-BEC stattering length $a_\mathrm{i}$ and the BEC scattering
length $a$. One finds
\begin{equation}\label{Delta-new}
\Delta = \frac{\omega_\perp}{4} \frac{m_\mathrm{B}^2}{m_\mathrm{i}^2} \frac{a_\mathrm{i}^2}{ R_\mathrm{TF} a}
\end{equation}
Here we have $m_\mathrm{i}=2m_\Br m_\mathrm{S}/(m_\Br+m_\mathrm{S})$. 
Thus assuming a tight transversal confinement with $\omega_\perp=2\pi 
\times 10^4$ Hz, a large scattering length between impurities and
BEC $a_\mathrm{i}= 200$nm, a small scattering within the BEC
$a=5$ nm, a small trap with $R_\mathrm{TF}=20 \mu$m, and $m_\mathrm{S} \approx m_\Br $ one finds
a conditional frequency shift of $2\pi \times 10^3$ Hz. 

As shown in Appendix C
the result of equation (\ref{DeltaHmWW1D}) should be valid 
as long as the following 
conditions are fulfilled
\begin{gather}
  \frac{z_1 -z_2}{R_\mathrm{TF}} \gg \pi \sqrt{\frac{\zeta}{2}}\\
  \frac{\delta r}{R_\mathrm{TF}} \gg \max \left\{\sqrt{ \zeta},\frac{z_0}{R_\mathrm{TF}}\right\} \,.
\end{gather}
Here, $\delta r$ denotes the distance of one of the impurities to the edge 
of the condensate and $\zeta=\hbar\omega_\Br/2\mu$ is the 
Thomas-Fermi parameter. Furthermore the interaction strength 
of the condensate has to fulfill the condition
\begin{equation}
  g_{\mathrm{1D}} \gg \frac{2 \hbar \omega_\Br R_\mathrm{TF}}{3 N_0} \,.
\end{equation}
Hence, we have the restriction
\begin{equation}
  \Delta \ll \min \left\{\omega_\Br, \frac{3 N_0 \kappa_\mathrm{1D}^2}{16\hbar^2 R^2_\mathrm{TF}\omega_\Br} \right\} \,.
\end{equation}


\section{Conclusions}


In the present paper we have analyzed the interaction of impurity 
atoms in a Bose-Einstein condensate localized at specific positions 
by tight confining potentials. It was shown that in addition to the 
level shift caused by s-wave scattering with the macroscopic condensate 
field there are also contributions from the interaction with 
vacuum fluctuations of the Bogoliubov phonons. The self- and conditonal
energy shifts were calculated for a BEC in a box with periodic
boundary conditions. It was shown that size and sign of 
the conditional energy shift depends on the separation of the
impurities and is largest for a highly anisotropic condensate 
geometry and for small interactions within the condensate. With 
increasing interaction of the condensate atoms the spatial dependence
becomes less and less pronounced. Motivated by these findings the
level shift in a quasi one-dimensional harmonic trap was calculated.
In the Thomas-Fermi limit a rather simple analytic expression 
was obtained from a Bogoliubov approach. For small trap sizes a
conditional frequency shift in the range of several kHz seems feasible
which could be of interest for the implementation of a quantum phase gate.


\section*{Acknowledgement}


This work was supported by the Deutsche Forschungsgemeinschaft through
the SPP 1116 ``Interactions in ultracold atomic and molecular gases''. 
A.K. thanks the Studienstiftung des Deutschen Volkes for financial support.


\appendix
\section{Derivation of the equation of motion for the statistical operator}
\label{HerleitgBGl}

The total statistical operator of both the condensate and 
the impurities is denoted by $\hat\chi$. 
Its time evolution is then given by the Liouville-von Neumann 
equation $i \hbar \partial_t \hat \chi (t)=
\left[ \hat H, \hat \chi(t)\right]$, where 
$\hat H=\hat H_\Br +\hat H_\mathrm{S}+\hat H_\mathrm{int}$ 
is the Hamiltonian of the whole system, 
with $\hat H_\Br$ being the Hamiltonian of the condensate, $\hat H_\mathrm{S}$ 
that of the impurities and $\hat H_\mathrm{int}$ the interaction.
Changing into the interaction picture
yields
\begin{equation} \label{vonNeumann}
  i \hbar \partial_t \tilde \chi (t) = \left[\tilde H_\mathrm{int} (t),\tilde \chi (t) \right] \,.
\end{equation}
Formal integration and resubstitution leads to
\begin{equation} \label{alt10}
\begin{split}
  i\hbar \partial_t \tilde{\chi}(t) = &\left[\tilde H_\mathrm{int}(t),\tilde\chi(t_0)\right] \\
 & + \frac{1}{i\hbar}\int_{t_0}^t \diff  t' \left[\tilde H _\mathrm{int}(t),\left[\tilde H _\mathrm{int}( t'), \tilde\chi( t')\right]\right] \,.
\end{split}
\end{equation}
Here, $t_0$ is the time when the interaction starts. 
The statistical operator for the impurity atoms can be obtained 
by tracing out the condensate, 
{\it i.e.} $\tilde\varrho(t)=\mathrm{Tr}_\Br
\left[ \tilde \chi(t)\right]$. This yields
\begin{equation} \label{statop1}
\begin{split}
 & i\hbar \partial_t \tilde{\varrho}(t) = \mathrm{Tr}_{\mathrm{B}}\left(\left[\tilde H_\mathrm{int}(t),\tilde\chi(t_0)\right]\right) \\
 &\quad+ \frac{1}{i\hbar}\int_{t_0}^t \! \diff  t' \, \mathrm{Tr}_\mathrm{B} \left(\left[\tilde H_\mathrm{int}(t),\left[\tilde H_\mathrm{int}( t'), \tilde\chi( t')\right]\right] \right)\,.
\end{split}
\end{equation}
Following the standard approach  we assume that the influence of the 
impurity atoms on the condensate can be neglected and that 
the statistical operator of the whole system seperates as 
\begin{equation}
  \tilde\chi (t) = \tilde\varrho(t)\otimes\tilde\varrho _\mathrm{B} (t) + \tilde\chi_\mathrm{corr} (t) \approx \tilde\varrho(t)\otimes\tilde\varrho_\mathrm{B} (t_0)\,.
\end{equation}
Furthermore since we have incorporated the mean-field contribution to the 
free Hamiltonian of the impurities, the 
expectation value of the interaction Hamiltonian vanishes.
, {\it i.e.} $\mathrm{Tr}_\mathrm{B}\left(\tilde\varrho_\mathrm{B}(t_0) 
\tilde H_\mathrm{int}(t) \right) =0$. 
With these approximations we obtain
\begin{equation}\label{BGLstatopA6}
\begin{split}
  \partial_t \tilde\varrho(t) = &-\frac{1}{\hbar^2}\int_{t_0}^t 
\!\diff  t' \, \\
  & \cdot \mathrm{Tr}_\mathrm{B}\left(\left[\tilde H_\mathrm{int}(t), 
\left[\tilde H_\mathrm{int} ( t'), \tilde\varrho( t')\otimes
\tilde\varrho_\mathrm{B}(t_0)\right]\right]\right) \,.
\end{split}
\end{equation}
The interaction Hamiltonian in the interaction picture can be expressed
as
\begin{equation}
  \tilde H_\mathrm{int}(t)=\sum_{\alpha,\beta} 
|\alpha\beta,t\rangle\langle \alpha\beta,t|
 \left( \frac{\kappa_\alpha}{2} 
\tilde B_1 (t) + \frac{\kappa_\beta}{2} \tilde B_2 (t)\right) \,.
\end{equation}
where
\begin{equation}
|\alpha\beta,t\rangle\langle \alpha\beta,t|= 
\eh^{\frac{i}{\hbar}\left(\hat {H}_\mathrm{S}
+ \hat{H}_\mathrm{B}\right)t} 
\left|\alpha\beta\right\rangle\left\langle \alpha\beta\right| 
\eh^{-\frac{i}{\hbar}\left(\hat{H}_\mathrm{S}+\hat{H}_\mathrm{B}\right)t}. 
\end{equation}
Substituting this into eq. (\ref{BGLstatopA6}) yields
\begin{widetext}
\begin{equation} \label{Mastereq}
  \begin{split}
  \partial_t \tilde \varrho_{\alpha\beta,\gamma\delta} (t) = - \frac{1}
{4 \hbar^2} \int_{t_0}^t \! \diff t' \, \tilde \varrho_{\alpha\beta,
\gamma\delta}(t') 
  &\left(\left\langle \tilde B_1(t) \tilde B_1(t')\right\rangle \left
(\kappa^2_\alpha - \kappa_\alpha\kappa_\gamma \right) 
   +\left\langle \tilde B_1(t) \tilde B_2(t')\right\rangle 
\left(\kappa_\alpha\kappa_\beta - \kappa_\beta\kappa_\gamma \right) \right. \\
  &+\left\langle \tilde B_2(t) \tilde B_1(t')\right\rangle 
\left(\kappa_\alpha\kappa_\beta - \kappa_\alpha\kappa_\delta \right)
   +\left\langle \tilde B_2(t) \tilde B_2(t')\right\rangle 
\left(\kappa_\beta^2 - \kappa_\beta\kappa_\delta \right)        \\
  &+\left\langle \tilde B_1(t') \tilde B_1(t)\right\rangle 
\left(\kappa_\gamma^2 - \kappa_\alpha\kappa_\gamma \right)
   +\left\langle \tilde B_1(t') \tilde B_2(t)\right\rangle 
\left(\kappa_\gamma\kappa_\delta - \kappa_\beta\kappa_\gamma \right)  \\
  &\left.+\left\langle \tilde B_2(t') \tilde B_1(t)\right\rangle 
\left(\kappa_\gamma\kappa_\delta - \kappa_\alpha\kappa_\delta \right)
   +\left\langle \tilde B_2(t') \tilde B_2(t)\right\rangle
 \left(\kappa_\delta^2 - \kappa_\beta\kappa_\delta \right) \right) \,.
\end{split}
\end{equation}  
\end{widetext}%


\section{Bogoliubov theory}\label{BogoliubovTheorie}


In this appendix we briefly summarize the main results of the 
Bogoliubov approach. 
We start with the hamiltonian of the Bose gas in $s$-wave-scattering 
approximation
\begin{equation} \label{HamiltonGas}
\begin{split}
  \hat{H}_\mathrm{B} =& \int \!\diff \mathbf{r} \, \hat{\psi}^\dagger(\mathbf{r})\left(-\frac{\hbar^2}{2m_\Br}\Delta +V_{\mathrm{ext}}(\mathbf{r}) -\mu \right) \hat{\psi}(\mathbf{r}) \\
  &+ \frac{g}{2} \int \!  \diff \mathbf{r} \, \hat{\psi}^\dagger(\mathbf{r})\hat{\psi}^\dagger(\mathbf{r}) \hat{\psi}(\mathbf{r})\hat{\psi}(\mathbf{r})  \,.
\end{split}
\end{equation}
The field operator $\hat \psi$ of the condensate is then devided
 into a $\mathbb{C}$-number function $\psi_0$ which represents 
the condensed part of the Bose-gas and an operator $\hat \xi$ 
of quantum fluctuations: $\hat \psi(\rb)=\psi_0(\rb)+\hat\xi(\rb)$. 
The wavefunction of the condensate is given by the Gross-Pitaevskii equation
\begin{equation}  \label{GPE}
  \left( -\frac{\hbar^2}{2m_\Br}\Delta +V_{\mathrm{ext}}(\mathbf{r}) 
-\mu + g \left| \psi_0(\mathbf{r})\right|^2 \right) \psi_0 (\mathbf{r}) = 0 \,.
\end{equation}
By plugging this into the Hamiltonian and neglecting terms of 
the order $\mathcal{O}(\hat\xi^3)$ and higher one gets
\begin{equation} \label{hamlin1}
  \begin{split}
      \hat H_{\mathrm{B}} \approx  H_{\mathrm{B}}^0 + 
\int \! &\diff \mathbf{r}  \,  \biggl\{ \hat \xi ^\dagger (\mathbf{r}) 
\left( -\frac{\hbar^2}{2m_\Br}\Delta +V_\mathrm{ext}(\mathbf{r}) -\mu\right) 
\hat \xi (\mathbf{r})  \\
      & + \frac{g}{2} \left( 4 \left|\psi_0 (\mathbf{r})\right|^2 \hat 
\xi ^\dagger (\mathbf{r})\hat \xi (\mathbf{r})\right. \\
      & \left.+ \psi_0^2 (\mathbf{r})\hat \xi  ^\dagger (\mathbf{r}) 
\hat \xi^\dagger (\mathbf{r}) + \left.\psi_0^{*}\right.^2 (\mathbf{r}) 
\hat \xi (\mathbf{r}) \hat \xi (\mathbf{r})\right)  \biggr\} \,.
  \end{split}
\end{equation}
The terms linear in $\hat \xi$ vanish because of the Gross-Pitaevskii 
equation. The term $H_\Br^0$ does not depend on operators and is 
without consequence.
In order to diagonalize the Hamiltonian we employ the Bogoliubov ansatz
\begin{gather} \label{Ansatzxi}
  \hat \xi (\mathbf{r}) = \left.\sum_{\nu} \right.^\prime u_\nu 
(\mathbf{r}) \hat b_\nu - v_\nu^* (\mathbf{r}) \hat b_\nu^\dagger \\
  \hat \xi^\dagger (\mathbf{r}) = \left.\sum_{\nu} \right.' u_\nu^* 
(\mathbf{r}) \hat b_\nu^\dagger - v_\nu (\mathbf{r}) \hat b_\nu \,.
\end{gather}
Here, $\hat b_\nu^\dagger$ and $\hat b_\nu$ are bosonic creation and 
anihilation operators of the Bogoliubov quasi-particles. 
The prime at the sum indicates that the ground state is excluded 
in the summation. If the wave functions $u_\nu$ and $v_\nu$ fulfill 
the Bogoliubov-de Gennes equations ($\psi_0$ is taken to be real)
\begin{gather} \label{BdG1}
   \left[-\frac{\hbar^2\Delta}{2m_\Br} +V_\mathrm{ext}(\mathbf{r}) 
-\mu\right] u_\nu  + g \left|\psi_0\right|^2 \left( 2u_\nu -v_\nu 
\right) = E_\nu u_\nu \\
   \left[-\frac{\hbar^2\Delta}{2m_\Br} +V_\mathrm{ext}(\mathbf{r}) 
-\mu\right] v_\nu + g \left|\psi_0\right|^2 \left( 2v_\nu -u_\nu 
\right) = -E_\nu v_\nu \,, \label{BdG2}
\end{gather}
with the normalization
\begin{gather} \label{BdGnorm1}
  \int \left\{ u_\nu(\mathbf{r}) u^*_{\nu'}(\mathbf{r})  
- v_\nu(\mathbf{r}) v^*_{\nu'}(\mathbf{r})\right\} \diff \mathbf{r} 
=\delta_{\nu \nu'}\\ \label{BdGnorm2}
  \int \left\{ v_\nu(\mathbf{r}) u_{\nu'} (\mathbf{r}) - 
u_\nu (\mathbf{r})v_{\nu'} (\mathbf{r})\right\} \diff \mathbf{r} =0 \,,
\end{gather}
the Hamiltonian takes the very simple form
\begin{equation} \label{Hamiltonbogol}
  \hat H_\mathrm{B}=H^0_\mathrm{B} - \left.\sum_\nu\right.' E_\nu \int \left| v_\nu(\mathbf{r}) \right|^2 \diff \mathbf{r} + \left.\sum_\nu \right.'E_\nu \hat b_\nu^\dagger \hat b _\nu \,.
\end{equation} 

With this the operators $\tilde \xi$ in the interaction picture 
can easily be calculated
\begin{equation}
   \tilde \xi (\mathbf{r},t) = \left.\sum_{\nu} \right.^\prime 
u_\nu (\mathbf{r}) \hat b_\nu \eh^{-iE_\nu t/\hbar} 
- v_\nu^* (\mathbf{r}) \hat b_\nu^\dagger \eh^{+iE_\nu t/\hbar} \,.
\end{equation}


\section{Validity of eq. (\ref{DeltaHmWW1D})}\label{Gueltigkeit}


In order to estimate the range of validity of the expression for
the conditional shift in TF approximation, eq. (\ref{DeltaHmWW1D}),
we start with the expression (see also eq.(\ref{Abkkorr})
\begin{equation}
\begin{split}
  \sum_{j=0}^M \frac{S_j(1,2)}{E_j} = &\sum_{j=0}^M  
\frac{1}{E_j}\int_{-R_{\mathrm{TF}}}^{R_{\mathrm{TF}}} \! 
\diff z \, \left| \phi_0 (z-z_1) \right|^2 \psi_0(z) f^-_j(z) \\
  & \times \int_{-R_{\mathrm{TF}}}^{R_{\mathrm{TF}}} \! 
\diff z' \left| \phi_0 (z'-z_2) \right|^2 \psi_0(z') f^-_j(z') \,,
\end{split}
\end{equation}
where $f^-_j=u_j-v_j$. 
By using (\ref{LoesungPetrov}) we find
\begin{equation} \label{Ausdruck0}
\begin{split}
  &\sum_{j=0}^M \frac{S_j(1,2)}{E_j} \sim  
\int_{-R_{\mathrm{TF}}}^{R_{\mathrm{TF}}} \! \diff z \, 
\int_{-R_{\mathrm{TF}}}^{R_{\mathrm{TF}}} \! \diff z' \, \\
  & \qquad \times \left| \phi_0 (z-z_1)\right|^2 \left| 
\phi_0 (z'-z_2)\right|^2 f_\mathrm{P}^M\left(\frac{z}{R_\mathrm{TF}}, 
\frac{z'}{R_\mathrm{TF}}\right)
\end{split}
\end{equation}
5
where we have introduced
\begin{equation}\label{fM}
  f_\mathrm{P}^M (x,x')=\sum_{n=0}^M \frac{2n+1}{2}P_n(x)P_n(x') \,.
\end{equation}
If $M\rightarrow \infty$ the sum approaches the $\delta$-function 
and we obtain equation (\ref{HmWW1Dfast0}). 
On the other hand the 
solutions (\ref{LoesungPetrov}) of the Bogoliubov-de Genne equations 
used here are only valid for \cite{Oehberg-PRA-1997}
\begin{equation} \label{BedingungOehberg1D}
  \frac{\delta r}{R_\mathrm{TF}}\gg \max\left[\frac{\sqrt{M(M+1)} \zeta}
{\sqrt{2}} ,\sqrt{\frac{\sqrt{2} \zeta}{\sqrt{M(M+1)}}} \, \right], 
\end{equation}
%
\begin{figure}[htb]
\begin{center}
  \scalebox{0.40}{\includegraphics{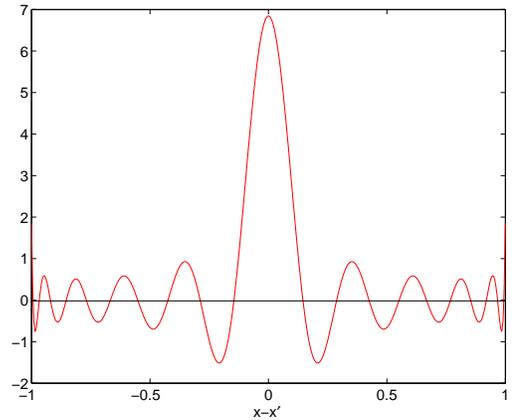}}
  \caption{Picture of $f_\mathrm{P}^{M}$ for $M=20$. 
\label{DeltaLegendrebild} }
\end{center}
\end{figure}%
%
%
where $\delta r$ is the distance from the edge of the condensate. 
This implies  $M \ll \sqrt{2} 
\delta r/R_{\mathrm{TF}}\zeta$ and with 
$\delta r \gg R_\mathrm{TF} \sqrt{\zeta}$, also following from
eq.(\ref{BedingungOehberg1D}) we arrive at 
$ M \ll \sqrt{\frac{2}{\zeta}}$.
Thus the limit $M \rightarrow \infty$ cannot be taken in
(\ref{fM}).
Nevertheless even for a finite but sufficiently large upper limit
of summation $M$ the sum is to a good approximation zero as can be seen as
follows: In figure \ref{DeltaLegendrebild} $f_\mathrm{P}^{20}$ is shown. 
One recognizes a pronounced central maximum. 
The first integral over $z$ in equation (\ref{Ausdruck0}) only contributes
if there is an overlapp of the maximum of $f_\mathrm{P}^M$ and the ground 
state $\phi_0(z-z_1)$. The same holds for the second integral over $z'$ 
and $\phi_0(z'-z_2)$. Hence, equation (\ref{Ausdruck0}) vanishes if 
the distance of the impurities is much bigger than the width of the 
central maximum.  
We thus need to estimate the width of this central peak. With the Stirling 
formula one finds asymptotically for large (and even) $M$
\begin{equation}
  f^{M}_\mathrm{P}(0,0) \approx \frac{M}{\pi} \,.
\end{equation} 
Since $\int f^{M}_\mathrm{P}(0,s)\,\diff s=1$ the width of the 
central peak can be approximated as $\Delta s=\pi/M$. 
This finally yields the condition
\begin{equation}
\frac{z_1-z_2}{R_\mathrm{TF}} \gg \frac{\pi}{M} \gg \pi \sqrt{\frac{\zeta}{2}}
\end{equation}
for which the sum in (\ref{Ausdruck0}) is approximately 0.
It should be noted that we have assumed the Thomas-Fermi 
limit $\zeta \ll 1$, which is essential for the analytic solution
of the Gross-Pitaevskii and Bogoliubov-de Gennes equations.

\newpage


\end{document}